\documentclass[10pt,conference]{IEEEtran}

\usepackage{graphicx}
\usepackage{amsmath}

\title{Contextual Information Retrieval based on Algorithmic Information Theory and Statistical Outlier Detection}

\author{\authorblockN{Rafael Martinez\authorrefmark{1},
Manuel Cebri\'an\authorrefmark{1}\authorrefmark{2},
Francisco de Borja Rodriguez\authorrefmark{1} and
David Camacho\authorrefmark{1}}
\authorblockA{\authorrefmark{1}Departamento de Ingenier\'ia Inform\'atica, Universidad Auton\'oma de Madrid\\
Rafael.Martinez@estudiante.uam.es, \{Manuel.Cebrian, F.Rodriguez, David.Camacho\}@uam.es\\
\authorrefmark{2}Department of Computer Science, Brown University,
mcebrian@cs.brown.edu}
}

\begin{document}
\maketitle

\begin{abstract}

The main contribution of this paper is to design an Information
Retrieval (IR) technique based on Algorithmic Information Theory 
(using the Normalized Compression Distance-NCD),
statistical techniques (outliers), and novel organization of data base structure. The paper shows how they can be
integrated to retrieve information from generic databases using long
(text-based) queries. 
Two important problems are analyzed in the
paper. 
On the one hand, how to detect ``false positives'' when the
distance among the documents is very low and there is actual similarity. On the other hand, we propose a way
to structure a document database which similarities distance
estimation depends on the length of the selected text. Finally,
the experimental evaluations that have been carried out to study
previous problems are shown.

\end{abstract}

\section{Introduction}

Any Information Retrieval (IR) system uses a query, and a document
collection, to search and retrieve from this collection the set of
most similar documents (usually ordered by any similarity
measure)~\cite{Baeza99,Doyle75,Rijsbergen79}. These kind of systems
are mainly designed to reduce information overload, and they allow
to access, find, and retrieve documents, video, music, or any type
of information that can be indexed in a database. The most popular
IR applications are the Web search engines such as Google, or Yahoo
search engines that allow to easily access to a huge available
amount of data.


Although it is quite hard to generalize, due the high number of
methods that have been designed to improve the accuracy of IR
systems, there are a high number of these systems that uses
classical IR theory as \textit{keyword} (usually named
\textit{terms} or \textit{descriptors}) matching techniques. These
are mainly based on a binary keyword matrix containing a vector of
words extracted from the set of documents to be retrieved. A zero
value means that one document has no correspondence with one
keyword, and non-zero value (usually one), indicates that this word
is a keyword for this document. Using this vector of features a
mathematical function calculates how much closer is the document to
the given query (see the Vector Space Model for a further
description~\cite{Salton75}).

Some interesting Web browsers that are not based on keywords
matching are Proximic (www.proximic.com), and Mercury News
(www.mercurynews.com). These browsers are based on a
\textit{context-based} searching over their documents, not just a
keyword search/match. Proximic uses a technology called ``pattern
proximity'' that uses a set of symbol sets as patterns to look for
for similarities in other documents. Therefore, Proximic does not
look at words, instead uses pattern recognition to understand the
``composition'' of text, not the text itself. Several advantages can
be considered from these kind of IR applications; it is possible to
use long queries (we are not restricted to use only some few key
words), it is possible to categorize the results, or to look for
information by its context.

There exist an important research in both, IR and Search engines
areas, related with knowledge, and files organization, automatic
text analysis, search strategies, automatic classification,
probabilistic retrieval, etc. Most of their techniques are
based on Natural Language, Data representation, machine learning,
pattern recognition,
and others techniques ~\cite{Smith76,Elkin78,Salton68,Robertson76}. However, in
the context of IR, the concept of information in a mathematical sense 
is not usually measured. In fact,
in many cases there no exist difference between both facts,
\textit{document} and \textit{information}.

In this paper a new approach to contextual-search information based
on Algorithmic Information Theory (AIT) techniques is presented. This approach uses
some elements from AIT theory such as Kolmogorov complexity estimation,
and other form statistics theory to compute the similarity between two
documents. Using techniques from both areas, we are trying to classify
documents by their content and structure (without using a set of given,
learned, or extracted keywords). In our approach the Kolmogorov Complexity of a document
is estimated using a compressor, an outliers are used to refine,
and improve, the accuracy of the retrieval process.

This paper is structured as follows. Section~\ref{ir} introduces
briefly some concepts about Kolmogorov Complexity (NCD) and
Statistics (outliers). Section~\ref{datarep} describes, and
analyzes, how to structure a set of documents that can be used by a
context-based search engine. Section~\ref{implementation} provides a
brief description of the IR application deployed to evaluate the proposed technique.
Section~\ref{experiments} provides the experimental results which
have been designed to obtain the best organization of the database,
and the optimal (statistical) parameters to maximize the accuracy of
the retrieval process. Finally, Section~\ref{conclusions} summarizes
the conclusions and describes the future lines of work.

\section{Information Retrieval based on Kolmogorov complexity and Statistics}
\label{ir}

The Kolmogorov Complexity of a text can be used to characterize the 
minimal amount of information needed to codify that particular text, 
regardless of any probability consideration. 
The Kolmogorov Complexity $K(x)$ of a string $x$, which is the size of the shortest program
able to output $x$ in a universal Turing machine, is an incomputable
problem too (due to the Halting problem), the most
usual (upper bound) estimation is based on data compression: the size of a compressed
version of a document $x$, which we will denote by $C(x)$ may be used as an estimation
of $K(x)$.

\subsection{Normalized Compression Distance}

A natural measure of similarity assumes that two objects $x$ and $y$
are similar if the basic \emph{blocks} of $x$  are in $y$  and
vice versa. If this happens we can describe the object $x$ by
referencing the blocks that belongs to $y$ , thus the description of
$x$  will be very simple using the description of $y$. This is
partially what a compressor does when catenates the $xy$ sequence:
a search for information shared by both sequences in order to reduce
the redundancy of the whole sequence. If the result is small, it
means that part of the information contained in $x$  can be used to
code $y$, following the similarity conditions described.

Previous, was formalized by Rudi Cilibrasi and Paul
Vitanyi~\cite{Cilibrasi05}, they
proved that it is possible to calculate an upper bound value of the
Kolmogorov complexity using compressors. This estimation, named the
Normalized Compression Distance (NCD), can be used as a similarity
distance between two objects. Therefore, the NCD distance, may be
used to cluster objects, or to sort them in a set of relevant
documents. The mathematical NCD formulation is shown in
equation~\ref{eq1}.

\begin{equation}
\label{eq1}
NCD(x,y)=\frac{\max\{C(xy)-C(x),C(yx)-C(y)\}}{\max\{C(x),C(y)\}},
\end{equation}

Where $C$ is an algorithm compression, $C(x)$ is the size of the
C-compressed version of $x$, and $C(xy)$ is the compressed size of
the catenation of $x$ and $y$. NCD generates a non-negative number
$0 \leq NCD(x,y)\leq1$. Distances near 0 indicate similarity between
objects, while those near 1 they shows dissimilarity. However, when
these methods are used it is necessary to make an analysis of the
document representation, the compressors have their own
characteristics and it is necessary to study how the documents will
be structured, and organized, to obtain a correct NCD distance
between two blocks of texts.

As the
quality of the NCD measure depends on the size of the objects
compressed (some well-known compressors do not correctly work if the object size exceeds some
memory constraints \cite{Cebrian05}), we make use of a windowless version of the classical Lempel-Ziv compression
algorithm.

\subsection{The Notion of Outlier}

The result of performing similarity calculations is usually a large
(quadratic in the number of submission in the corpus) number of
pairwise distances, a number of which will be very low in case of
similarity occurrence. The goal of an information retrieval tool is to
flag these distances as the ones connecting relevant documents. 

When based only on the bulk of distance values, the process
of evaluation relevant candidates depends on the determination of a
similarity threshold value. If a pair of documents have a similarity
distances which falls below this threshold, it will be marked for
inspection. Otherwise, the pair will not be marked. 

However, the task of locating a good initial threshold has received
little attention in the literature, and is therefore left completely to
the personal decision of the search engine designer. 

Under certain conditions, it is possible to quantify the \emph{amount of surprise} 
presented by a distance value within a sample of distances. This problem
is completely equivalent to the one of finding outliers in a data sample, a classical
problem in statistical data analysis. Quoting Barnet and
Lewis\cite{barnett} 
\begin{quote}
We shall define an outlier in a set of
data to be an observation (or subset of observations) which appears
to be inconsistent with the remainder of that set of data.
\end{quote}

The term `surprisingly
small distances' is thus equivalent to
the term lower outliers as used in statistics. In other words, the
set of distances which should be marked as connecting relevant documents 
in a query corresponds to what a reasonable
statistical test will consider as lower outliers within the full set
of distances. As far as the authors are aware, this is the very
first application of outlier detection techniques to Information Retrieval,
despite its natural analogy.

\section{Data Knowledge structure and organization}
\label{datarep}

Database research~\cite{date75introduction} is a
hot topic related with a wide number of topics like; Web database
systems, information integration, data mining, parallel database
systems, scientific databases, visualization of large data sets,
database performance,\ldots The research in this area (usually
related with others such as Artificial Intelligence, Software
Engineering or Physics for instance) tries to solve different
problems (i.e. high performance and reliability accessing the data)
for any particular domain (Web, networks, huge data sets,\ldots).

This section analyzes the problem of data structure, and
organization, from our (technique) perspective. The target is to
obtain a reliable data organization that allows to optimize both,
find the best block of inside a particular document (this block, or
document elemental unit, that better fit with the query), to
minimize the time necessary to find the most similar documents.

\subsection{Database organization}

Like in any database repository, we have a set of documents that
need to be stored and indexed to allow their future retrieval. Any
document in our database will be divided in elemental units (named
blocks) from 1 Kb to $N$Kb. The maximum size ($N$) of a block
depends with the compressor used, because any standard compressor
uses a window which defines the best behavior in the algorithm (for
instance the maximum window for LZ-compressor is 32 Kb).

\begin{figure}[htp]
\centering
\includegraphics[width=3 in]{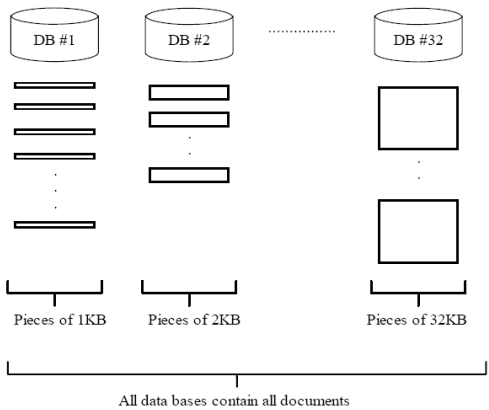}
\caption{Distributed Database organization}\label{databases}
\end{figure}

This is a critical design aspect in our retrieval method, because if
we use arbitrary length blocks, the NCD estimation of these blocks
will not fit correctly (because the compression algorithm it is not
able to find correctly the similarities) with the user query.
Figure~\ref{databases} shows an schematic representation of a
distributed database build by 32 data repositories where each
document has been divided in blocks from 1Kb to 32Kb, therefore it
will be used a redundant distributed database, where each database
can use their indexes to retrieve this elemental units.

Another problem that needs to be analyzed is related with the user
query. Although any document can be pre-processed off-line (we know
its size once it has been stored in the database), it is not
possible to know what will be the size of the user query. With our
approach the user query, may vary from a sentence, or a paragraph,
to a complete file. The variable length of the user query will be
handle as our previous files, so any user query will be processed
into elemental units from 1Kb to $N$Kb, if the size of the user
query is greater than $N$ KB, it will be processed into $N$Kb blocks
(as any other file). This user query processing will aid to obtain a
more accurate NCD value.

\subsection{File structure elemental division}

If we consider a file like a sequence of characters (i.e. string) we
can divide it into blocks of approximately 1024 bytes, 2048
bytes, etc, until the complete division of the file. These blocks
build the elemental units of a particular file, that finally are
indexed and stored in the corresponding database. However, the
results, in the retrieval process, of this structure organization
could not work so well at it would be expected.

The problem is newly related with the base technique used
(compression) to look for a particular document. Any compressor is
an algorithm designed to detect several similarities, or
regularities (structural, statistical,\ldots) in the documents. What
   happens if a document is truncated in basic block? Possibly the compressor will not be able to find
their similarities, and therefore our similarity distance will not
be so good as we desire.

\begin{figure}[htp]
\centering
\includegraphics[width=3 in]{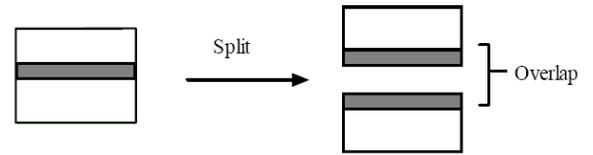}
\caption{Overlapping allows maintaining some structural data using
redundant information}\label{overlap}
\end{figure}

To smooth previous problem our file structure division uses an
\emph{overlap} in the blocks (see Figure~\ref{overlap}). Overlapping
means that any block ($N_i$) contains some information (bytes) of
the precedent block ($N_{i+1}$). This allows to preserve some of the
structure that could be contained in a particular block, on the
other hand it is a high memory cost technique so it is necessary to
evaluate carefully. This overlap is a configuration parameter in our
search engine that can be set from 1\% to 99\% (i.e. if 10\% is
selected a 40Kb file needs, using 10Kb blocks, about 50Kb of
memory).

\section{Search engine}
\label{implementation}

A simple search engine prototype has been designed to allow make
different experimental evaluations of the proposed technique. This
search engine uses a set of graphical interfaces to allow:
preprocessing a set of document repositories and store them into our
database organization; deploy these databases in the search engine;
calculate the NCD for each stored document; show the set of
documents found from a particular user query (with the NCD distance
for each block); show the documents found, and highlight those
blocks (inside a particular document) with the best similarity.

\begin{figure}[htp]
\centering
\includegraphics[width=3in]{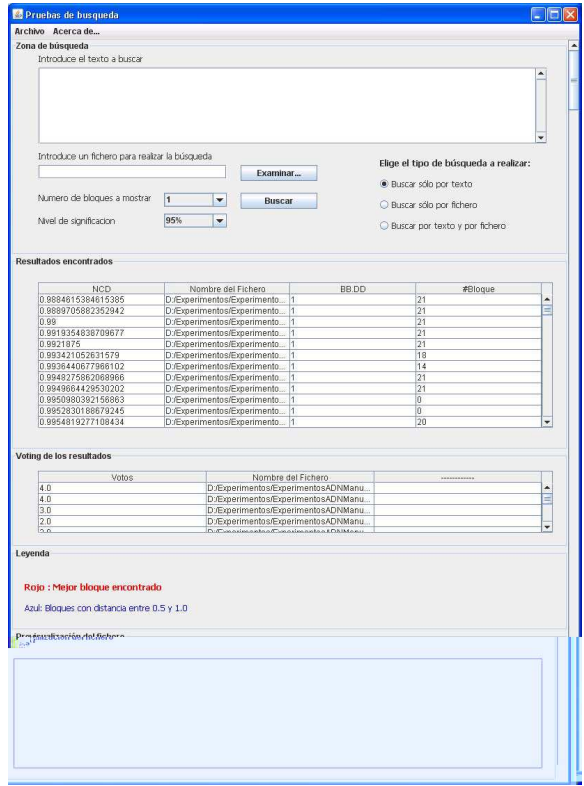}
\caption{Search engine user interface}\label{gui}
\end{figure}

As Figure~\ref{gui} shows, the main GUI of the search engine is
built by four work areas. The first one is a text area where the
user can enter some text to be queried, or introduce any existing
file with the content that wants to search for. This are also
provides some customizable parameters like the number of blocks to
be shown, or the limit value for outliers search.

The second area shows those blocks that have been found. This area
provides information about the NCD value of the block, the file
name, the database where it has been stored, and its relative
position position in the whole file (i.e. database 1KB, block 21).

Once all the blocks have been obtained, the third area provides the
results of the total voting for each file/blocks. The files are
ordered by increasing similarity. Any file can be selected to show
the complete file, and the blocks that have been found, in the last
work area (bottom area in the gui).

\subsection{The Voting Method}
Each time we have a query we compute the NCD distance from the query
to several elements  (document fragments) of the DB, those whose
size is in the same size interval of the  query, the 2 following
intervals and the previous interval (see point 3.3). In this way we
obtain several sets of distances, $d_1, d_2, \ldots$

We have gathered empirical evidence that distances from a query to
DB elements (document fragments) which are not relevant are normally
(gaussian) distributed (data not shown).

Additionally, we have experimental evidence that distances from a
query to relevant elements are generally lower outliers of  the
distribution, i.e. distances which are abnormally low as to have
been generated by a  normal distribution which some particular mean
and standard deviation.

Put it differently, the probability of a non-relevant distance being
as low as a relevant one is extremely small. Therefore, it is fair enough
 to propose the following querying method:

 \begin{enumerate}
   \item Choose some size interval $S$.

   \item Compute the distances from the query the elements in the database
   which are in the size interval $S$. We assume the majority of the distances
   are non-relevant and some very few distances are relevant. Therefore the
   statistical distribution of the data will be normal (gaussian).

   \item Find outliers in this distance set. Each one of the fragments whose
   distance to the query is an outlier counts as one vote to the document it belongs to.

   \item Go to 1, until no more intervals can be selected.

 \end{enumerate}

At the end of this loop we sum the votes from several size
intervals, and we highlight those fragments which received a vote.


\subsection{Finding outliers}
The Hampel identifier, is maybe the most extensively studied (see
[23]) outlier-finding method for normally distributed data sets. 
Additionally, there is empirical data proving that it
outperforms other tests in many
applications\cite{Pearson, Wilcox}.

The Hampel identifier works as follows. Let $X(1), X(2), \ldots ,X(N)$ be the ordered distances
$X_1, X_2, X_3 ,X_N$. Let $M$ be the median of the sample, and $S$ be the median absolute deviation
form the median of the sample.

The Hampel identifier, adapted to lower outliers, is a rule which identifies as relevant
all distances of the sample $X$ satisfying:

\begin{equation}
(M - X)/S > g(N;\alpha)
\end{equation}

where the function $g(N;\alpha)$ serves for standardizing the identifier in the
following way (see \cite{Davies-Gather}, p. 783, standardization (4)):

\begin{equation}
\Pr \left (\text{no outliers in sample}\right)=\Pr \left (\frac{|X_{(N)}-M|}{S}<g(N;\alpha) \right)=1-\alpha
\end{equation}

The function $g(N; \alpha)$ does not have an analytic form and we estimate by means 
of a Montecarlo simulation.

This method can handle a large number of outliers, and is resistant to the problems
appreciated in non-parametric approaches \cite{AC}. The designer may choose which threshold to use,
being $\alpha=0,01$ and $\alpha=0,05$ the most usual ones.

\section{Experimental Results}
\label{experiments}

Our experimental setup consists of a database with a total amount of 100 peer-reviewed scientific publications 
by 17 different authors, all of them belonging to the Department of Computer Science at Universidad Aut\'onoma de Madrid, and therefore in the Computer Science field.

As explained in section IV.A, we use voting for considering a relevant/irrelevant result. A non-zero value 
of voting is enough to consider the document containing the voted fragment as relevant. Although
a higher number of votes represents a higher relevance, this is not taken into account in 
this experimental validation (but is used to rank results by their decreasing number of votes).



Our first set of experiments is intended as a proof-of-concept, and consists of selecting an abstract
from one of the documents included in the database, and using it as a query. 
The desired output, \emph{true positive}, is therefore the document
from which the abstract was extracted. On the other hand, a \emph{false positive} is any other document.







We consider our search engine as a binary classifier system with two classes, relevant and irrelevant. One can see the results using a \emph{ROC curve}, a graphical plot of the \emph{sensitivity} (rate of true positives) vs. (1 - \emph{specificity}) (rate of false positives) as $\alpha$ (our discrimination threshold) goes from $0$ (total discrimination, no documents retrieved) to $1$ (no discrimination, all documents are retrieved).

\begin{figure}[htp]
\centering
\includegraphics[width=3.5 in]{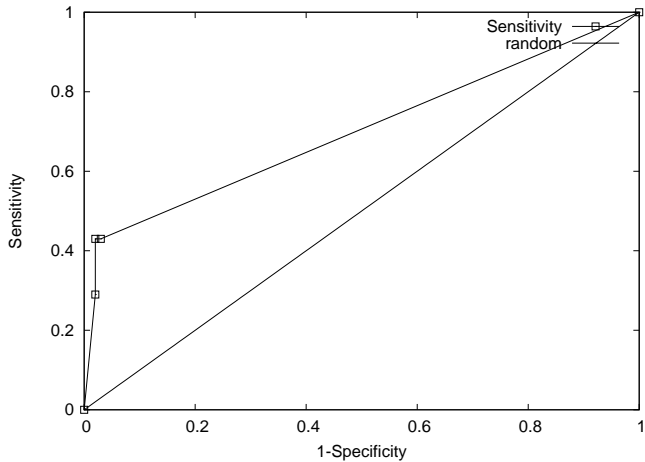}
\caption{Representative ROC curve for a query of the experiment \#1 kind.}\label{fig:exp1gr3}
\end{figure}

Figure~\ref{fig:exp1gr3} depicts a representative query result of the above described kind of experiments. We also depict the ROC curve of a random binary classifier for the sake of comparison. Results above the random curve represent positive evidence of information retrieval, and the faster the curve separates from the random curve, the better the search engine performs.

\begin{figure}[htp]
\centering
\includegraphics[width=3.5 in]{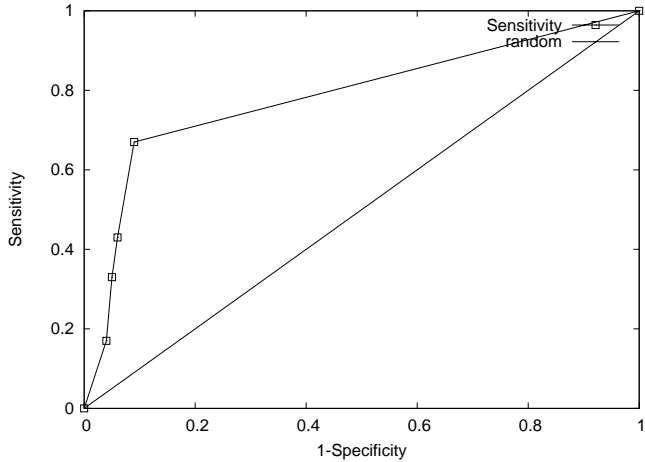}
\caption{Representative ROC curve for a query of the experiment \#2 kind.}
\label{fig:exp2gr4}
\end{figure}

In a second step we remove the abstract from every document of the database, and we repeat the previous queries. The true positive and false positive consideration is unchanged. A representative result is depicted in Figure~\ref{fig:exp2gr4}.








\begin{figure}[htp]
\centering
\includegraphics[width=3.5 in]{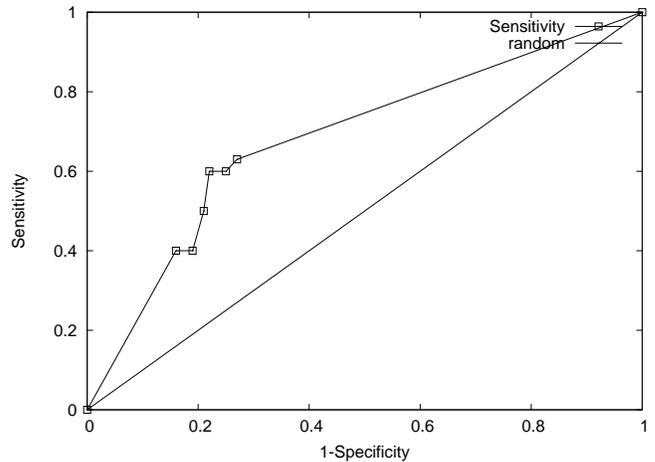}
\caption{Representative ROC curve for a query of the experiment \#3 kind.}\label{fig:exp3gr3}
\end{figure}

In the final step, we choose 20 documents which scientific classification subject coincides with one or more subjects
of the documents in the database. This is done using the \emph{SpringerLink} search engine (www.springerlink.com).
We then select 5 fragments from each document, and use each of them as a query to the database. True positive results are those documents whose subject coincides with the query subject, and false positive are those which do not. A representative result of single query is shown in figure \ref{fig:exp3gr3}.


%





We therefore have a clear-cut experimental evidence of positive information retrieval in
experiments ranging from simple literal text-coincidence queries to subject-affine document exploration.


\section{Conclusions and future work}
\label{conclusions}

This paper has described a new IR technique based on Algorithmic Information
Theory and Statistical techniques. The Information Theory technique is based on the utilization
of the Normalized Compression Distance. This distance has been used as a feasible approximation
of the Normalized Information Distance (based on the ideal notion of Kolmogorov Complexity). 
For a particular document, this measure is used to compare how close, or similar,
two documents are, without using keywords, or any other traditional IR technique.

The paper has described the organization of the database and the files structure, both essential
characteristics to implement the proposed technique. It has been shown how the detection of outliers
is used to obtain a better precision in the retrieving process in a natural way.

Several experiments have been carried out with to main objectives. On one hand to fit some
parameters (i.e overlapping in the databases, Hampel identifier threshold and others) of the search engine, and in the other hand
to test in a set of scientific documents, how is working the method. The encouraging efficacy of the 
search engine together with its great simplicity and generality makes it a promising line of research
towards alternative information retrieval systems.

In the future other database repositories will be used to study this
technique (i.e. genetic databases and classical text books). Other compression algorithms, like
PPMZ, BZIP2 or GZIP, will be integrated in the search engine to evaluate how these algorithms
affects to the retrieval technique. Finally, other techniques, like intelligent
distortion in textual documents that have been designed to obtain better complexity estimations
 (in form of most accurate NCD distances)~\cite{granados08},
could be used to improve the IR technique proposed.

\section*{Acknowledgments}
This work was supported by TIN 2004-04363-CO03-03, TIN 2007-65989,
CAM S-SEM-0255-2006, S-0505/TIC/000267 and TSI 2005-08255-C07-06.

\bibliographystyle{IEEEtran}
\bibliography{cites}

\begin{thebibliography}{10}
\providecommand{\url}[1]{#1}
\csname url@rmstyle\endcsname
\providecommand{\newblock}{\relax}
\providecommand{\bibinfo}[2]{#2}
\providecommand\BIBentrySTDinterwordspacing{\spaceskip=0pt\relax}
\providecommand\BIBentryALTinterwordstretchfactor{4}
\providecommand\BIBentryALTinterwordspacing{\spaceskip=\fontdimen2\font plus
\BIBentryALTinterwordstretchfactor\fontdimen3\font minus
  \fontdimen4\font\relax}
\providecommand\BIBforeignlanguage[2]{{%
\expandafter\ifx\csname l@#1\endcsname\relax
\typeout{** WARNING: IEEEtran.bst: No hyphenation pattern has been}%
\typeout{** loaded for the language `#1'. Using the pattern for}%
\typeout{** the default language instead.}%
\else
\language=\csname l@#1\endcsname
\fi
#2}}

\bibitem{Baeza99}
R.~Baeza-Yates and B.~Ribeiro-Neto, \emph{Modern Information Retrieval}.\hskip
  1em plus 0.5em minus 0.4em\relax Addison-Wesley, 1999.

\bibitem{Doyle75}
L.~Doyle, \emph{Information Retrieval and Processing}.\hskip 1em plus 0.5em
  minus 0.4em\relax Melville Publishing Co., 1975.

\bibitem{Rijsbergen79}
C.~van Rijsbergen, \emph{Information Retrieval}.\hskip 1em plus 0.5em minus
  0.4em\relax Butterworths, 1979.

\bibitem{Salton75}
G.~Salton, A.~Wong, and C.~S. Yang, ``A vector space model for automatic
  indexing,'' \emph{Communications of the ACM}, vol.~18, no.~11, p.
  613â620, 1975.

\bibitem{Smith76}
L.~Smith, ``Artificial intelligence in information retrieval systems,''
  \emph{Information Processing and Management}, vol.~12, pp. 189--222, 1976.

\bibitem{Elkin78}
N.~Elkin, ``Information concepts for information science,'' \emph{Journal of
  Documentation}, vol.~34, pp. 55--85, 1978.

\bibitem{Salton68}
\emph{Automatic Information Organization and Retrieval}.\hskip 1em plus 0.5em
  minus 0.4em\relax McGraw-Hill, 1968.

\bibitem{Robertson76}
S.~Robertson and K.~Jones, ``Relevance weighting of search terms,''
  \emph{Journal of the American Society for Information Science}, vol.~27, pp.
  129--146, 1976.

\bibitem{Cilibrasi05}
R.~Cilibrasi and P.~Vitanyi, ``Clustering by compression,'' \emph{IEEE
  Transactions on Information Theory}, vol.~51, no.~4, pp. 1523--1545, 2005.

\bibitem{Cebrian05}
M.~Cebri\'an, M.~Alfonseca, and A.~Ortega, ``Common pitfalls using normalized
  compression distance: what to watch out for in a compressor,''
  \emph{Communications in Information and Systems}, vol.~5, no.~4, pp.
  367--384, 2005.

\bibitem{barnett}
V.~Barnett and T.~Lewis, ``Outliers in statistical data,'' \emph{Wiley New
  York}, 1994.

\bibitem{date75introduction}
\BIBentryALTinterwordspacing
C.~J. Date, \emph{An Introduction to Data Base Systems}.\hskip 1em plus 0.5em
  minus 0.4em\relax Reading,: Addison-Wesley, 1975. [Online]. Available:
  \url{citeseer.ist.psu.edu/anywehere.html}
\BIBentrySTDinterwordspacing

\bibitem{Pearson}
R.~Pearson., ``Outliers in process modeling and identification,'' \emph{Control
  Systems Technology}, vol.~27, pp. 10(1):55--63, 2002.

\bibitem{Wilcox}
R.~Wilcox, \emph{Applying contemporary statistical techniques}, 2003.

\bibitem{Davies-Gather}
L.~Davies and U.~Gather, ``The identification of multiple outliers,''
  \emph{Journal of the American Statistical Association}, vol.~88, no. 423, pp.
  782--792, 1993.

\bibitem{AC}
\BIBentryALTinterwordspacing
M.~Freire, M.~Cebri\'an, and E.~del Rosal, ``Ac: An integrated source code
  plagiarism detection environment,'' 2007. [Online]. Available:
  \url{arXiv:cs/0703136.}
\BIBentrySTDinterwordspacing

\bibitem{granados08}
A.~Granados, M.~Cebri\'an, D.~Camacho, and F.~de~Borja~Rodriguez, ``Evaluating
  the impact of information replacement on normalized compression
  distance-driven text clustering,'' in \emph{IEEE Information Theory
  Workshop}, (Submitted) 2008.

\end{thebibliography}

\end{document}